\documentclass[twocolumn,10pt]{asme2e}
\usepackage{graphicx}
\usepackage{subcaption}
\usepackage{tabularx}
\usepackage{amsmath}
\usepackage{url}

\confshortname{IDETC/CIE 2025}
\conffullname{the ASME 2025 International Design Engineering Technical Conferences \&\\
              Computers and Information in Engineering Conference}

\confdate{August 17-August 20}
\confyear{2025}
\confcity{Anaheim}
\confcountry{California, USA}

\title{Co-optimize condenser water temperature and cooling tower fan using high-fidelity synthetic data}

\author{Gulai Shen
    \affiliation{
	AI Research \\
	Nantum AI\\
	New York, NY 10021\\
    }	
}

\author{Gurpreet Singh
    \affiliation{	AI Research \\
	Nantum AI\\
	New York, NY 10021\\
    }
}

\author{Ali Mehmani\thanks{Corresponding Author: 	amehmani@prescriptivedata.io}   
   \affiliation{	AI Research \\
	Nantum AI\\
	New York, NY 10021\\
    }
}

\begin{document}

\maketitle    

\begin{abstract}
This paper introduces a novel method for optimizing HVAC systems in buildings by integrating a high-fidelity physics-based simulation model with machine learning and measured data. The method enables a real-time building advisory system that provides optimized settings for condenser water loop operation, assisting building operators in decision-making. The building and its HVAC system are first modeled using eQuest. Synthetic data are then generated by running the simulation multiple times. The data are then processed, cleaned, and used to train the machine learning model. The machine learning model enables real-time optimization of the condenser water loop using particle swarm optimization. The results deliver both a real-time online optimizer and an offline operation look-up table, providing optimized condenser water temperature settings and the optimal number of cooling tower fans at a given cooling load. Potential savings are calculated by comparing measured data from two summer months with the energy costs the building would have experienced under optimized settings. Adaptive model refinement is applied to further improve accuracy and effectiveness by utilizing available measured data. The method bridges the gap between simulation and real-time control. It has the potential to be applied to other building systems, including the chilled water loop, heating systems, ventilation systems, and other related processes. Combining physics models, data models, and measured data also enables performance analysis, tracking, and retrofit recommendations.
\end{abstract}

\section*{INTRODUCTION}
According to the U.S. Energy Information Administration (EIA), buildings account for about 40\% of energy consumption, 75\% of electricity use, and 35\% of all carbon emissions in the US \cite{buildingenergy, shoemaker_nrel_2023}. Through improved evaluation, modeling, and control, buildings can achieve greater energy efficiency, comfort, and carbon neutrality \cite{shen2025PeopleHour, Mehmani2018DEBES}. 

\subsection*{Building thermal and energy modeling}
Depending on the available knowledge and data for different environments, three main types of model are commonly used in building energy modeling: black-box, grey-box (surrogate), and white-box models. The advantages and limitations of each method are discussed in the following sections. Researchers have explored ways to compare different models or combine them to address specific problems and enhance performance. Neto and Fiorelli \cite{Neto2008compare} compared the building modeled in EnergyPlus \cite{energyplus} as a white-box model and an artificial neural network (ANN) as a black-box model. Their results showed a 10\% error for the ANN model and a 13\% error for the EnergyPlus model.

Black-box models rely purely on data to construct a model, typically using regression or neural networks based on collected data. The availability, quality, and distribution of the gathered data can significantly impact the accuracy of the model. This approach allows for the development of accurate models without detailed information about the building and removes the process of building and calibrating the physics-based building models. However, extensive high-quality data may not always be available, especially for new buildings or obsolete buildings lacking recorded data. Further, the distribution of the data might only align with the typical operational patterns of the building. For unseen scenarios, the model performance might degrade significantly. Additionally, the parameters obtained may not have clear physical meanings. Mohandes et al. \cite{Mohandes2019ANN} provide a comprehensive review of existing AI approaches for building energy modeling and optimization. Yang et al. \cite{Yang2005ANN} compare an adaptive ANN with a static ANN for building energy prediction, achieving a coefficient of variance of approximately 0.2. This method is also applicable to different types of buildings. Biswas et al. \cite{Biswas2016ANN} demonstrate the use of ANN for energy prediction in residential buildings.

Grey-box models have limited knowledge of the actual physical environment but simplify it into resistor-capacitor (RC) models to simulate the thermal behavior of buildings. Different configurations of resistors and capacitors are incorporated into the system. The common choices include models 2R2C, 3R2C, and 6R2C. With a reasonable choice of the simplified model and tuning of the model parameters, a realistic model can be achieved. Danza et al. \cite{Danza2016model} construct an RC model capable of achieving an overall error of 2.7\% between the simulated and measured data. Berthou et al. \cite{Berthou2014model} also develop multiple RC models with the best performance of less than 2\% error for the fitted data 84\%. The advantage of the grey-box is the physical meaning it carries and the reduction in the amount of data needed to produce the model. For energy model and simulation, surrogate models \cite{Ali2016AMR} share a similar idea of simplifying complete physics into a few principle components that best describe the system to predict energy consumption. However, both methods could be difficult to maintain the balance between simplicity and performance, as well as to determine the dominant parameters of the systems.

White-box models include detailed physics-based equations and all the main simulation software and tools which use all the available information of the building and external conditions to create a simulated environment as close to reality as possible. Popular platforms include EnergyPlus, TRNSYS, eQuest, Modelica, and others. White-box models have been the way people previously predicted energy consumption and increased efficiency for the past several decades. They can produce good models, but they are time-consuming to make and require computational power. For buildings and equipment that have missing blueprints, have undergone retrofits and modifications, or have been running for extensive years without data measuring, the accuracy of the model could be severely affected.

\subsection*{Model-based optimizations}
Model-based building optimization is achieved based on one of the models generated above. The model predictive control (MPC) problem is solved using optimization techniques, including genetic algorithm (GA), particle swarm optimization (PSO), gradient-based algorithm, and many others. The focus for different problems varies from optimized energy, optimized cost, optimized set-points, optimized operation sequence, etc. Lu et al. (2003) \cite{Lu2003CWopt} built an empirical model with physical constraints that leads to optimized condenser water loop operations with assumptions. Cutillas et al. \cite{Cutillas2017opt} optimize the cooling tower considering both electricity and water consumption. Liu et al. \cite{Liu2018opt} use TRNSYS with measurements to generate an optimized control strategy that increases energy efficiency more than two times. One of the key challenges to incorporating high-fidelity models into real-time control includes computational cost, time, and interface to existing building management and automation systems.

Demand-side management (DSM) \cite{Mehmani2019DSM, shen2024DSM_GHG} is an important aspect of building optimization consideration and can be achieved by combining the efforts of building designers, operators, and occupants. DSM can be formulated as linear optimization problems to reduce energy cost with a combination of tariff models \cite{Kepplinger2015DSM}. DSM can also have different objectives or even multi-objectives in one problem formulation \cite{Paull2010DSM}. Some of the main goals of DSM include reducing energy consumption, peak demand, and cost, as well as adjusting and regulating frequency and voltage. DSM can even be further considered to a group of end-users rather than just one for a more robust system in all\cite{Mohsenian2010DSM, shen2021data}.

\section*{METHODOLOGY}

\subsection*{Problem formulation}
\begin{figure}
\centering
\includegraphics[width=0.9\linewidth]{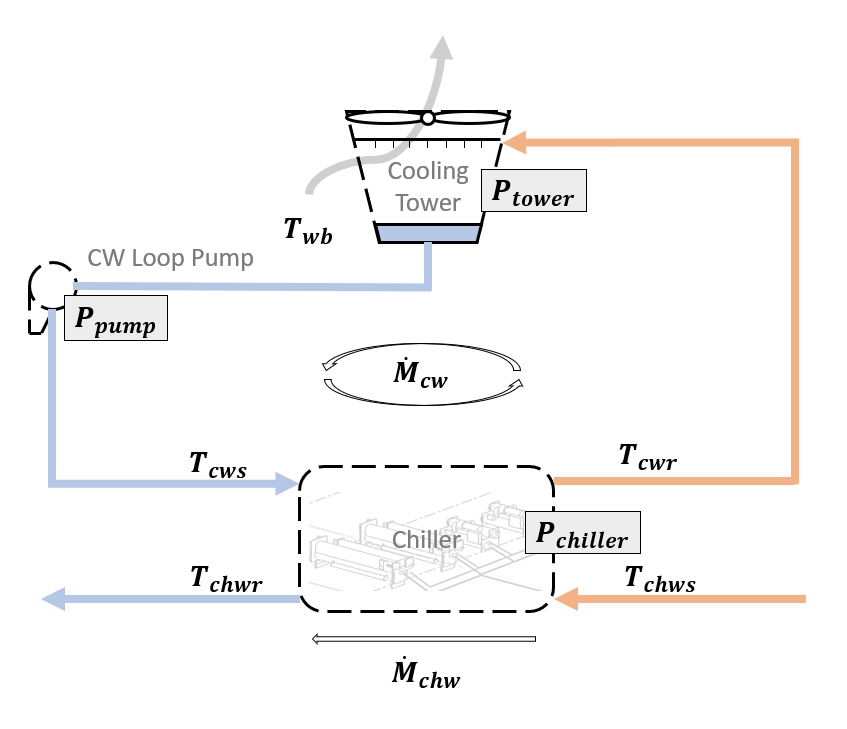}
\caption{\uppercase{Simplified HVAC System with the condenser water cooled by cooling tower.}}
\label{fig:simpHVAC}
\end{figure}

This paper focuses on the optimization of the condenser water loop, including cooling towers, condenser water pumps, and chillers, shown as a simplified HVAC system in Figure \ref{fig:simpHVAC}. The water in this circulation is pumped through the condenser, raising the temperature to take the heat load of the building and the work performed by the chiller compressor. The heat is rejected through the cooling tower, leading to the cooled water, which is pumped back into the chiller. Zhang et al \cite{Zhang2018TRNSYSopt} provides an example of building a function from data between the cooling load and the equipment power and tries to find a balance where total energy is minimized.

\begin{figure*}
\centering
\includegraphics[width=0.9\linewidth]{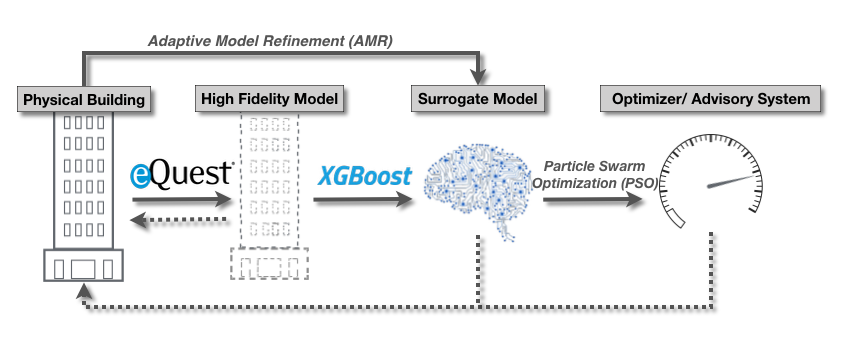}
\caption{\uppercase{Employed method procedure}}
\label{fig:procedure}
\end{figure*}

Depending on the system configurations, a fully optimized condenser water loop runs in an optimal combination of chillers, pumps, and cooling towers with optimal speed and temperature settings that minimize the total energy consumption or total energy cost incurred while adequately coping with the load of the building.

The procedure for this paper shown in Figure \ref{fig:procedure} includes:\
\begin{enumerate}
  \item Building an eQuest model of the building with exhaustive detail for the condenser water loop
  \item Adjusting performance curves for components to better mimic the behavior of the actual system
  \item Gathering data by adjusting different variables, including condenser water temperature, weather, load, occupancy schedule, etc., to prepare enough data.
  \item Physics analysis, data analysis, correlation analysis, and model variable determining.
  \item Training models that capture the behavior of the system and can be run in real-time.
  \item Develop an optimizer that uses the trained model to produce real-time optimal setting suggestions.
  \item Improving the model with recursive interactions between trained model, simulation, and real data. 
\end{enumerate}

\subsection*{Synthetic data generation with eQuest}
eQuest is a simulation tool based on DOE-2, which allows parametric analysis and graphical design and analysis of the building envelope and simulation results \cite{equest}. It is a well-established, sophisticated, and continuously improving graphical interface for the DOE-2 simulation engine. Its capability to provide meaningful insights to building engineers has been proven by many. Ke et al. \cite{Ke2013equest} show an example of how to tune eQuest parameters to get a high-fidelity model and how to do parametric analysis to show the effects of different options on building energy. Because of the algorithm and assumptions made, different simulators may produce varying results even with the same tuning, and the error of simulation might be significant in some specific cases. Zhu et al. \cite{Zhu2012equest} note that eQuest cannot accurately model the multiple zones with significantly varying conditions. Further, Singh et al.\cite{Singh2012equest} show that even with the same calibration of the building envelope, different software will produce different results within a reasonable range.

\begin{figure*}
\centering
\includegraphics[width=0.9\linewidth]{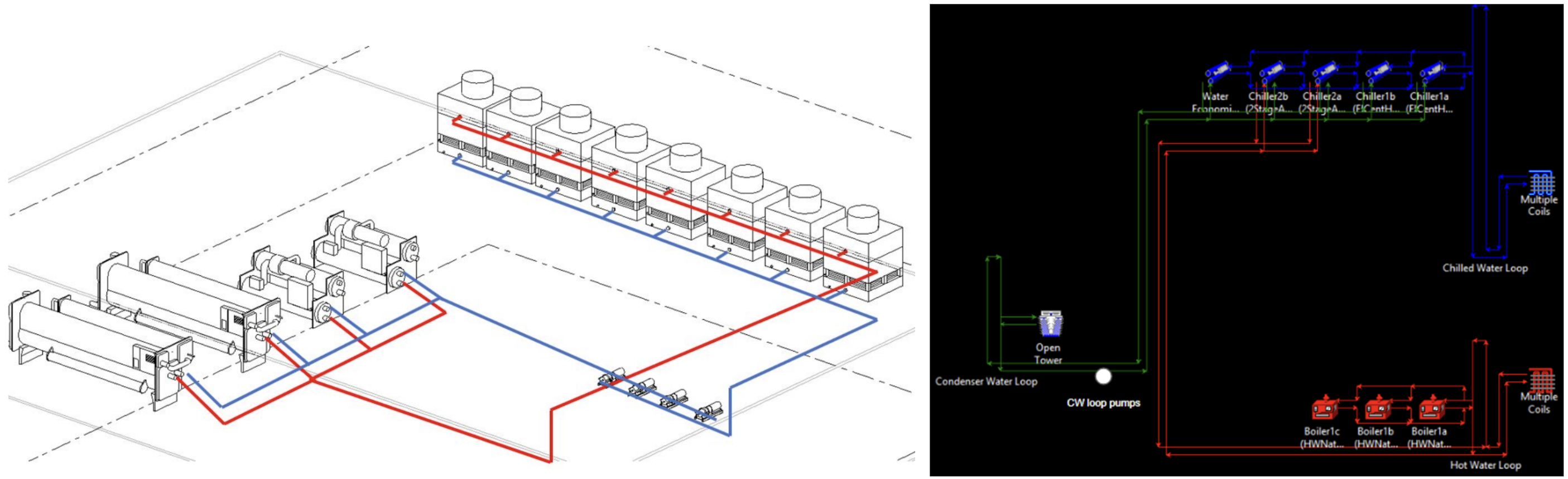}
\caption{\uppercase{Condenser water loop model setup in eQuest and in BIM software}}
\label{fig:cw_loop_model}
\end{figure*}

\begin{figure}
\centering
\includegraphics[width=0.9\linewidth]{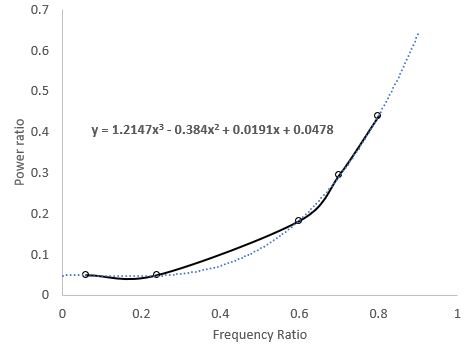}
\caption{\uppercase{Fitted fan curve}}
\label{fig:fancurve}
\end{figure}

In this project, an eQuest model of a commercial building in NYC is built. The model is simplified in its architecture, air units, and zone dividing but is fine-tuned for the condenser water loop shown in Figure \ref{fig:cw_loop_model}. Due to the current configuration of the HVAC system, the 34-story high building is cooled by two 1350-ton electric chillers, an 8-cell cooling tower, and constant-volume condenser water loop pumps. Thus, optimal operation would have a condenser water supply setting which minimizes the energy consumption of the condenser water loop with the combination of cooling tower fans, chillers, and pumps. To capture the behavior of the condenser water loop more precisely, data for the chiller and cooling tower are measured to tune the curves, which are imported into the eQuest simulation. Figure \ref{fig:fancurve} shows the fitted fan curve used in the simulation. To generate data with varying cooling load, weather, occupancy, condenser water temperature, and number of fans that result in different power consumption of the chiller and cooling tower, multiple simulation runs are performed during the cooling months. 

\subsection*{Model dominant variable determination}
\subsubsection*{Chiller model}

\begin{figure}
\centering
\includegraphics[width=0.9\linewidth]{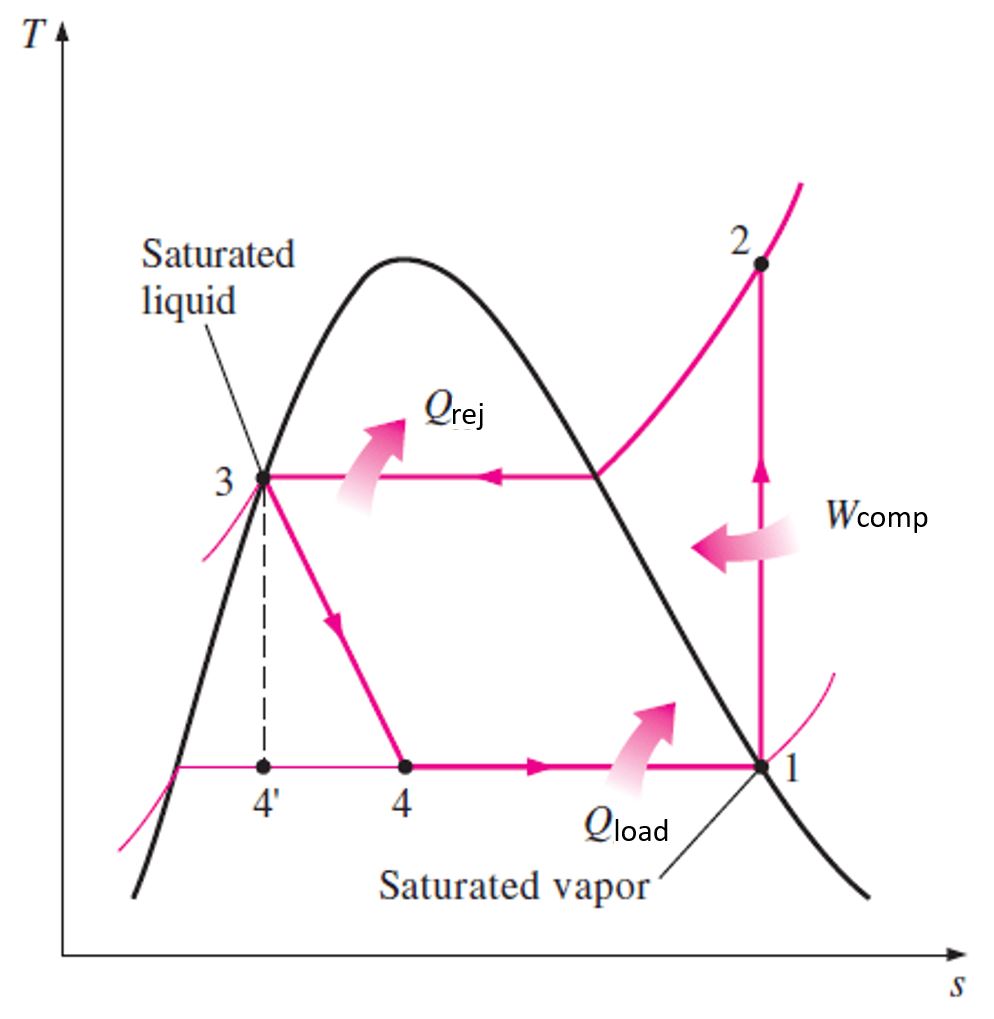}
\caption{\uppercase{Refrigeration cycle of a chiller \cite{chiller_cycle_2014}}}
\label{fig:chiller_cycle}
\end{figure}

Chiller can be modeled as a refrigerant cycle shown with the compressor, the single most dominant power consumer of a chiller. The work done by the compressor is:

\begin{align}
    \dot W_{comp} &= \dot m (h_2 - h_1) \\
    P_{chiller} &= \frac{\dot W_{comp}}{\eta} \\
    \dot Q_{rej} &= \dot m (h_2 - h_3) = \dot M_{cw} c_p (T_{cwr} - T_{cws}) \\
    \dot Q_{load} &= \dot m (h_1 - h_4) = \dot M_{chw} c_p (T_{chwr} - T_{chws})
\end{align}

Where $\dot m$ is the mass flow rate of the refrigerant, $h_2$ is the enthalpy at state 2, and $h_1$ is the enthalpy at state 1. $P_{chiller}$ is the power of the chiller based on $\eta$ which is the efficiency of the compressor at current condition. $Q_{rej}$ is the amount of heat rejection which can be calculated from the temperature of the chiller water supply temperature$T_{chws}$, return temperature$T_{chwr}$ and the flow rate$\dot M_{chw}$, $Q_{load}$ is the cooling load of the building that can be calculated from the condenser water supply temperature$T_{cws}$, return temperature$T_{cwr}$ and flow rate$\dot M_{cw}$. Thus, both the load and the rejection side conditions could influence the chiller power consumption. It is reasonable to assume that $P_{chiller}$ can be experimentally modeled as a function of $T_{cws}$ and $Q_{load}$. The data correlation will be checked in later sections.

\subsubsection*{Cooling tower model}
The main source of heat rejection comes from the evaporation of water. Thus, the power consumption of the fan would largely depend on the ambient air wet-bulb temperature, the water basin temperature it sets to drop to, and the total amount of heat rejection required. Jin et. al \cite{Jin2007CT} provided a simplified model of a mechanical draft cooling tower using both physics equations and data. Serna et. al \cite{Serna2010CT} formulated a Mixed Integer Nonlinear Programming problem to optimize the cooling tower. Both showed the importance of the effects of ambient temperature and water temperature on the performance and power consumption of the cooling tower. To further prove the assumption, the data correlation will again be checked in later sections.

\subsection*{Model training with XGBoost}
"Extreme Gradient Boost (XGBoost) is an implementation of gradient-boosted decision trees designed for speed and performance." \cite{xgboost} XGBoost suits tabular data very well and suits this specific problem, which has a limited range of possible values. In total, 6 models are trained in this example, which predicts the chiller power, heat rejection, and cooling tower power under 2, 4, 6, and 8 fans running in parallel.

\subsection*{Optimization with particle swarm optimization}
To find the optimized value for the condenser water temperature set point and the number of fans running, this paper uses the Particle Swarm Optimization (PSO) algorithm (1995) \cite{kennedyr1995pso}. PSO algorithms have been widely used in MPC to improve thermal comfort \cite{wang2024IEW_index}, minimize energy consumption, and reduce the life cycle cost of equipment systems \cite{Xu2005opt, Yang2012opt}. Since two variables in this problem need to be optimized, and one of them is an integer value, a modification is made to the way the particles are generated and updated to be a Mixed Integer Particle Swarm Optimization. The condenser water temperature setting is a continuous variable within a certain range, while the number of fans is a discrete value. Thus, particles are generated uniformly across the range of the temperature set points while covers all the possible fan numbers. Through iterations, the variable representing the number of fans is not updated, ensuring the cover of all possible cases of a number of fans while iterating. The temperature, with other variables, is, however, updated based on the global best result and the particle best result available with the equations shown below:

\begin{equation}
    V_i(k+1) = w*V_i(k)+c_1*(Best_i-X_i)+c2(Best_{all}-X_i)
\end{equation}
Where V is the velocity, i is the index of the particles, k is the iteration number, $Best_i$ is the best-ever position for each particle, $Best_global$ is the best-ever position for all the particles, X is the position of the particle, and $w$, $c_1$, and $c_2$ are the constant coefficients. This equation updates the velocity of each particle, i.e., how each particle will be moving in the next time step. Then, the position of each particle is updated by:
\begin{equation}
    X_i(k+1) = X_i(k)+V_i(k+1)
\end{equation}

\subsection*{Time of use(ToU) tariff structure}
\begin{figure*}
\centering
\includegraphics[width=0.9\linewidth]{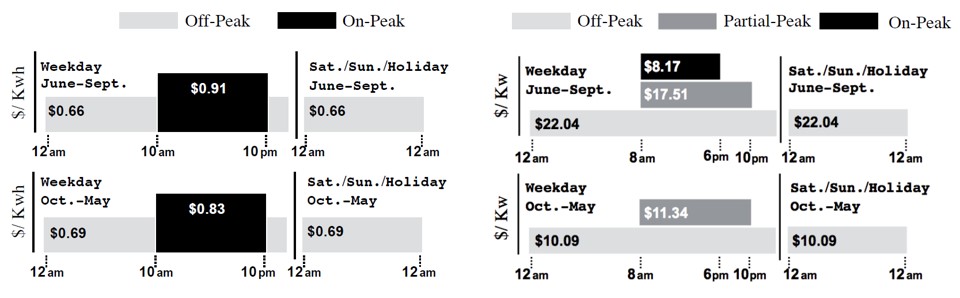}
\caption{\uppercase{SC9-Rate III ToU demand-based tariff from ConEdison}}
\label{fig:elecbill}
\end{figure*}

With both the optimizer and the measured data over June and July, an analysis of the actual potential savings for the building is performed under the ToU demand-based tariff \cite{Ali2015model}. The tariff based on Consolidated Edison Company of New York (ConEdison) \cite{conedison} is shown in Figure \ref{fig:elecbill}. Both total energy consumption and highest demand would impact the tariff. A model is built based on the schedule, demand, and consumption which is later used to compare the cost of operation under current settings and optimized settings with a close-to-realistic measure.

\section*{RESULTS AND DISCUSSION}

\subsection*{Simulation data}
\begin{figure*}[h]
\centering
\begin{subfigure}{0.45\textwidth}
    \centering
    \includegraphics[width=\textwidth]{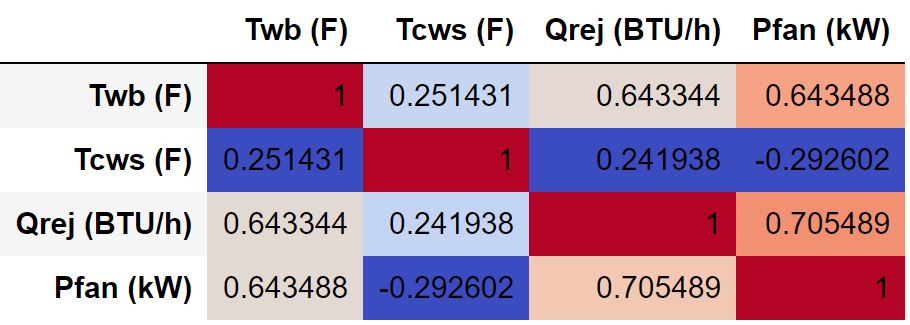}
    \caption{\uppercase{Fan correlation matrix}}
    \label{fig:fan_corr}
\end{subfigure}
\hfill
\begin{subfigure}{0.45\textwidth}
    \centering
    \includegraphics[width=\textwidth]{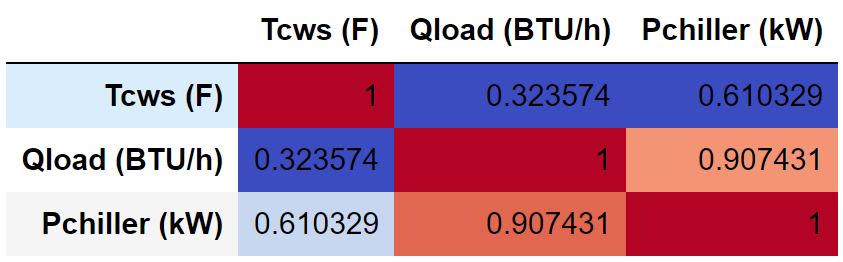}
    \caption{\uppercase{Chiller correlation matrix}}
    \label{fig:chiller_corr}
\end{subfigure}
\caption{\uppercase{Comparison of correlation matrices: (a) Fan correlation matrix, (b) Chiller correlation matrix}}
\label{fig:corrmatrix}
\end{figure*}

Each simulation run generates the hourly results over a whole simulation year using typical weather data in New York City. The output in each step includes data on outside wet bulb temperature, building load, condenser water supply and return temperatures, number of fans operating, cooling tower fan power consumption, and chiller power consumption. To check the dominant variables for the power consumption of the fan and the chiller, the correlation matrix is shown in Figure \ref{fig:corrmatrix}. It is observable that wet-bulb temperature($T_{wb}$) and the amount of heat rejection have a strong correlation with the power of the fan($P_{fan}$). Meanwhile, both the condenser water supply temperature($T_{cws}$) and cooling load on the building($Q_{load}$) have a strong correlation with the chiller power. These numbers confirm the assumptions and simplification made to build a surrogate model based on the high-fidelity simulation.

\begin{table}[h]
\centering
\renewcommand{\arraystretch}{1.2} 
\setlength{\tabcolsep}{3pt} 
\begin{tabular}{|c|c|c|c|c|c|}
\hline
\textbf{Twb (°F)} & 65 & 66 & 67 & 68 & 69 \\
\hline
\textbf{Avg. Diff. (\%)} & 10.34 & 7.74 & 4.6 & 0.11 & 8.01 \\
\hline
\textbf{Twb (°F)} & 70 & 71 & 72 & 73 &- \\
\hline
\textbf{Avg. Diff. (\%)} & 2.5 & 0.48 & 11.21 & 5.76 &- \\
\hline
\end{tabular}
\caption{\uppercase{Average percentage difference at various wet-bulb temperatures.}}
\label{tab:twb_diff}
\end{table}

\subsection*{Model and optimizer}
The models generated allow users to change the variables and see the impact on the change in energy consumption. Before moving to optimization, both the eQuest model and the machine learning model generated are verified with the real building data measured in June and July. First, even though the building envelope and the air-side system were not modeled in detail, the eQuest building load prediction is still capable of generating data with a maximum 11\% and average 5.64\% error in the summer months with a wet bulb temperature from 65 to 73 degrees Fahrenheit shown in Table \ref{tab:twb_diff}. 


Furthermore, the chiller model is also compared to the measured data. In general, the model captures the tendency of the measured data and produces a mean bias error of less than 1\%. Nevertheless, measured data are much noisier and oscillate above and below the model prediction a lot, causing the coefficient of variation of the mean error to be around 20\%. It is important to increase the accuracy further in the future, either by using a better simulation model or by using the measured data to refine the ML model. Lastly, the optimizer using particle swarm optimization further allows users to understand the optimal settings under the current building cooling load and wet-bulb temperature.

\begin{table}[h]
\centering
\renewcommand{\arraystretch}{1.2} 
\setlength{\tabcolsep}{3pt} 
\begin{tabular}{|c|c|c|c|c|}
\hline
Month & kWh Sav. (\$) & kWh Sav. (kWh) & Total Sav. (\$) & Sav. \% \\
\hline
June  & 1615  & 21230  & 1615  & 8\% \\
\hline
July  & 605   & 7960   & 609   & 3\% \\
\hline
\end{tabular}
\caption{\uppercase{Energy savings summary for typical summer months of June and July.}}
\label{tab:savings}
\end{table}

\subsection*{Saving analysis}
\begin{figure*}[h]
\centering
\begin{subfigure}{0.45\textwidth}
    \centering
    \includegraphics[width=\textwidth]{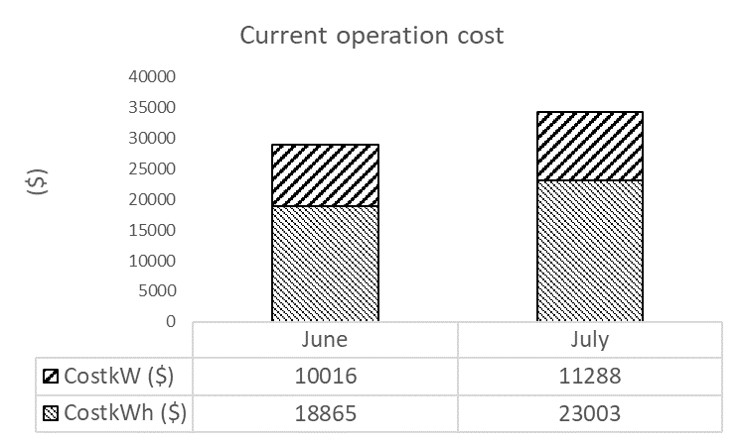}
    \caption{\uppercase{Current cost}}
    \label{fig:currentbill}
\end{subfigure}
\hfill
\begin{subfigure}{0.45\textwidth}
    \centering
    \includegraphics[width=\textwidth]{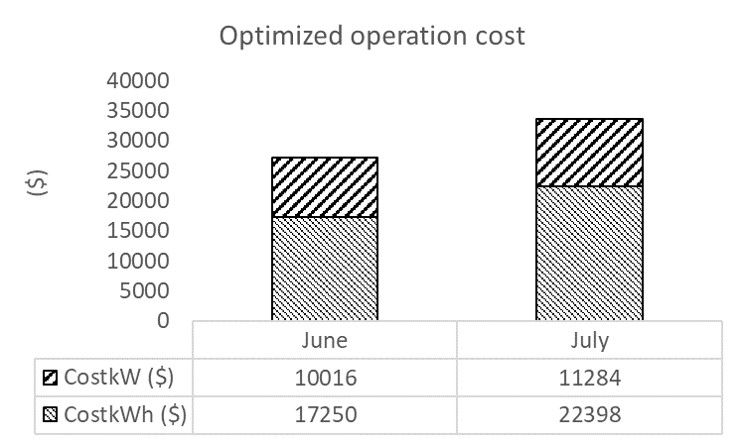}
    \caption{\uppercase{Optimized cost}}
    \label{fig:optimizedbill}
\end{subfigure}
\caption{\uppercase{Comparison of electricity costs: (a) Current cost, (b) Optimized cost}}
\label{fig:eleccost}
\end{figure*}

The saving analysis is performed based on data from June and July with data at a 15-minute interval. The costs, including the electric cost for the chillers and cooling towers, are shown in Figure \ref{fig:eleccost} based on the demand-based tariff model introduced before. They are generated by running the model at the measured temperature settings and the optimized settings. The potential savings in dollars are around \$1,600 for June and around \$600 for July, as shown in Table \ref{tab:savings}. It is observed that the kW costs based on the maximum power are almost the same before and after optimization, while the kWh costs have an obvious drop. It is reasonable that the existing control systems are generally designed to be near-optimal in the worst-case scenario. Further, the maximum cooling load at a high temperature left fewer choices for building operators to set the system, which allows the system to be running at a closer to optimal point. Nevertheless, there does exist great potential in kWh savings and a cooler month like June tends to generate more savings than a hotter month of July. 

\subsection*{Operation look-up table}
At the same time, for the optimizer to be able to run online to provide building operators with suggestions about cooling tower fan operation and condenser water temperature setting, an offline look-up table can be constructed ahead for certain buildings under certain limit conditions. It can also be used as a reference in the starting stage of testing and verifying the validity of the model and optimizer. 
 
\begin{figure*}
\centering
\includegraphics[width=0.95\linewidth]{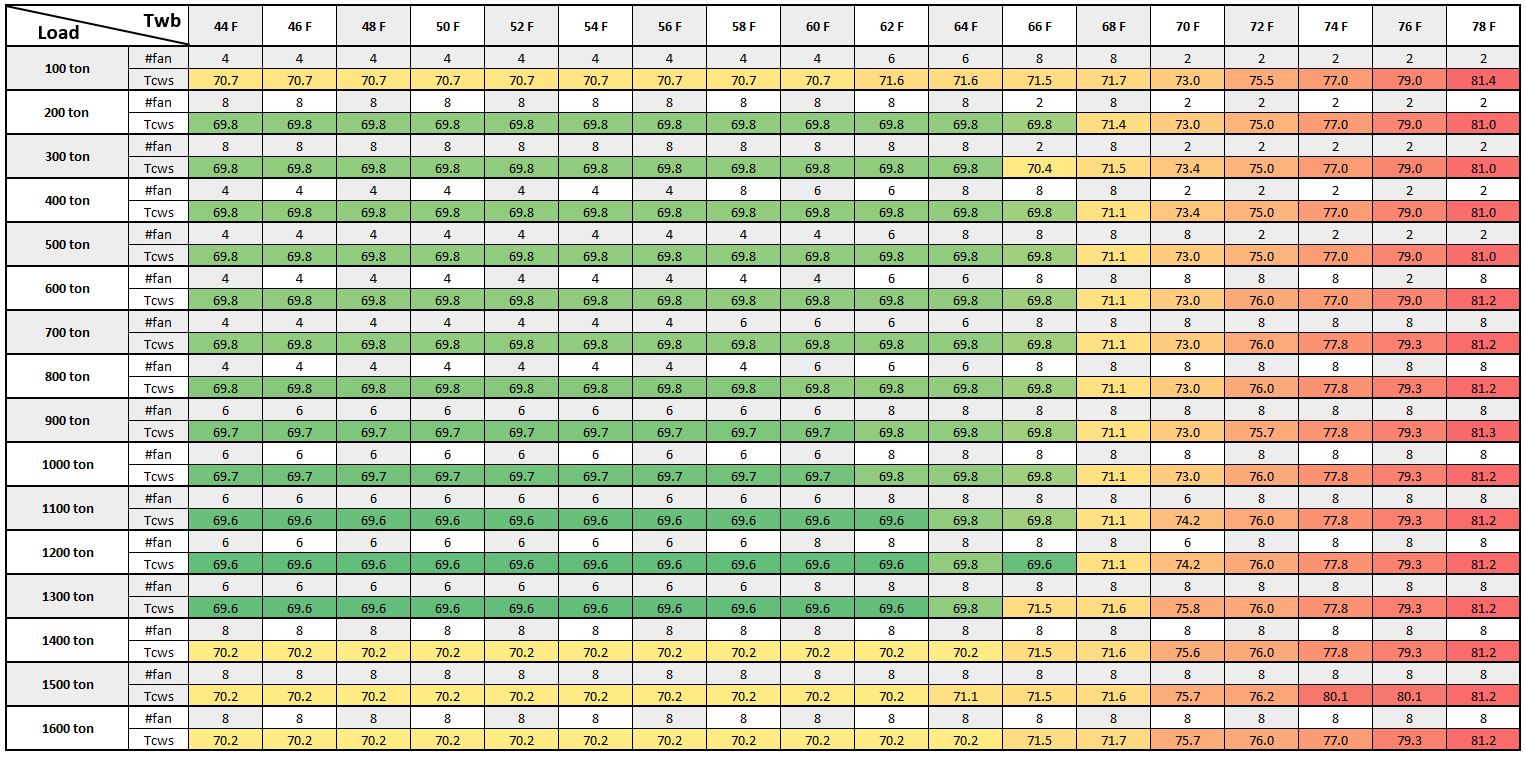}
\caption{\uppercase{Operation table at different load conditions for operating with 2 electric chiller}}
\label{fig:optable}
\end{figure*}

The table provides building engineers and operators with optimal condenser water temperature and number of fan settings for cooling towers under certain cooling load and outside wet-bulb temperature. The look-up table covers the range of expected loads and weather conditions. The table could be useful in the initial stage of testing the model and optimizer in an actual building environment and can be continuously updated in batches offline. An example of the table for a commercial building in New York City operating with 2 chillers and chilled water settings at 44 degrees Fahrenheit is shown in Figure \ref{fig:optable}.

\section*{CONCLUSION}

In conclusion, we developed and validated a novel method to co-optimize the condenser water supply temperature and cooling tower fan stages for the condenser water loop of the building HVAC system under the objective of minimizing energy consumption of the cooling tower fan, condenser water loop pumps, and chillers given the cooling load. Combining the physics-based model of simulation and data-based model with machine learning, we are able to solve the building energy optimization problem efficiently and implement it in real-time. The accuracy of the model for the building envelope is, on average, 5.64\%; the chiller model produces a mean bias error less than 1\% regardless of the coefficient of variation, which is around 20\%. The solution provides an online real-time optimizer and an offline operation look-up table that is readily available for building operators to follow. The savings calculated using real data measured in June and July show a potential of \$2,200 in total for two months. 
 
To further increase the accuracy of the model prediction, adaptive model refinement(AMR)\cite{ghassemi2020AMR} can be used to take measured data to tune a better model. A similar method can be applied to other large, complicated systems, such as the chilled water loop and the air-side system, as well as other operation cases, such as free cooling, heating, etc. Furthermore, with the combination of the models and data, scalable solutions can be possible across different types of building systems and buildings.

    

\begin{acknowledgment}

The authors would like to thank Gene Boniberger, John Gilbert, John J. Gilbert IV, Evan Torkos, Amit Paul, Lauren Long, and Gary Chance from Nantum AI \cite{nantumAI}, along with Rudin Management Company, for their support and contributions to this project.

\end{acknowledgment}

%
% ...


\bibliographystyle{asmems4}
\bibliography{references}



\end{document}